\documentclass[%
 reprint,
 amsmath,amssymb,
 aps,
 prl
]{revtex4-2}

\usepackage{graphicx}
\usepackage{dcolumn}
\usepackage{bm}
\usepackage{hyperref}

\usepackage{xcolor}

\usepackage{amsmath, amssymb}

\usepackage{comment}
\usepackage{braket}
\newcommand{\bbR}{{\mathbb R}}

\begin{document}
\preprint{APS/123-QED}

\title{Monte Carlo studies of the emergent spacetime
  in the polarized IKKT model}

\author{Chien-Yu C\textsc{hou}}
\email{ccy@post.kek.jp}

\author{Jun N\textsc{ishimura}}
\email{jnishi@post.kek.jp}

\author{Cheng-Tsung W\textsc{ang}}
\email{ctwang@post.kek.jp}

\affiliation{KEK Theory Center, Institute of Particle and Nuclear Studies,\
High Energy Accelerator Research Organization,\
1-1 Oho, Tsukuba, Ibaraki 305-0801, Japan}
\affiliation{Graduate Institute for Advanced Studies, SOKENDAI,\
1-1 Oho, Tsukuba, Ibaraki 305-0801, Japan}

\date{\today; preprint: KEK-TH-2740}

\begin{abstract}
  The IKKT matrix model has been investigated as a promising nonperturbative formulation
  of superstring theory.
  One of the recent developments concerning this model is
  the discovery
  of the dual
  supergravity solution corresponding to 
  the model obtained after supersymmetry-preserving mass deformation, which is dubbed
  the polarized IKKT model.
  Here we perform Monte Carlo simulations of this model in the case of matrix size $N=2$
  for a wide range of the deformation parameter $\Omega$.
  While we reproduce precisely the known result for the partition function obtained
  by the localization method developed for supersymmetric theories,
  we also
  calculate the observables, which were not accessible by previous work,
  in order to
  probe the spacetime structure emergent from the dominant matrix configurations.
  In particular, we find that the saddle point corresponding to the original IKKT model
  is smoothly connected to the saddle represented by the fuzzy sphere dominant at large $\Omega$,
  whereas the dominant configurations
  become diverging commuting matrices at small $\Omega$.
\end{abstract}

\maketitle


\textit{Introduction---}
The IKKT model (or the type IIB matrix model)~\cite{Ishibashi:1996xs,Aoki:1998bq}
was originally proposed as a matrix regularization of type IIB superstring theory,
which is conjectured to provide 
a constructive definition of second-quantized string theory
in the large-$N$ limit.
It takes the form of
the large-$N$ reduced model of
$10$-dimensional $\mathcal{N}=1$ super Yang–Mills
theory~\cite{Eguchi:1982nm,Parisi:1982gp,Gross:1982at,Gonzalez-Arroyo:1982hyq},
which can also be viewed as 
the effective theory of D-instantons~\cite{Witten:1995im,Green:1997tv}.
As a nonperturbative string-theoretic model without a predefined spacetime background,
it offers a promising approach to studying the dynamical emergence
of spacetime~\cite{Aoki:1998vn,Hotta:1998en}.
A key question, for example, is whether our $(3+1)$-dimensional
spacetime emerges
as a dominant eigenvalue distribution of the
matrices~\cite{Kim:2011cr,Kim:2011ts,Kim:2012mw,Steinacker:2017vqw,Steinacker:2017bhb,Sperling:2019xar,Hatakeyama:2019jyw,Steinacker:2021yxt,Brahma:2022ikl,Klinkhamer:2022frp,Hirasawa:2024dht}
from the underlying $(9+1)$-dimensional superstring theory.

One of the issues that has not been explored until recently in this model
is the gauge-gravity duality \cite{Maldacena:1997re,Aharony:1999ti},
which can be derived from superstring theory
by considering coinciding $N$ D$p$-branes as a background.
The low energy effective theory of the D$p$-branes is given by
$(p+1)$-dimensional U($N$) super Yang-Mills theory,
which can be thought of as a holographic description of the supergravity
solution in the large-$N$ and strong coupling limits.
The IKKT model corresponds to the extreme case $p=-1$,
where the super Yang–Mills theory reduces to a matrix integral
without a time coordinate unlike the BFSS model \cite{Banks:1996vh}, which corresponds to $p=0$.
An important consequence of this is that there is no time-derivative in the action,
which implies that the Yang-Mills coupling constant can be absorbed by rescaling
the matrices properly. 
The corresponding
supergravity solution has only  been
discussed based on the D-instanton charge~\cite{Gibbons:1995vg,Ooguri:1998pf,Ciceri:2025maa}.


Recently it has been realized that
these difficulties in the gauge-gravity duality for $p=-1$
can be overcome \cite{Hartnoll:2024csr,Komatsu:2024bop}
by considering the \emph{polarized IKKT matrix model},
which is obtained by applying
a SUSY-preserving deformation \cite{Bonelli:2002mb} to the original model.
The deformation introduces a mass scale $\Omega$, which
explicitly breaks the ${\rm SO}(10)$ symmetry down to
${\rm SO}(3) \times {\rm SO}(7)$,
and turns classical solutions into $\mathfrak{su}(2)$ representations.
Following a similar analysis \cite{Lin:2004nb,Lin:2005nh}
in the SUSY-deformed BFSS (or the BMN) matrix model~\cite{Berenstein:2002adc},
a family of supergravity solutions preserving 16 supercharges has been identified
as the holographic dual of the classical solutions in the polarized IKKT
model~\cite{Hartnoll:2024csr,Komatsu:2024bop}.

For $\Omega\gg1$, the path integral is dominated by
the classical solution that corresponds to
the $N$-dimensional irreducible representation of $\mathfrak{su}(2)$,
which may be viewed as the maximal
fuzzy sphere \cite{Azuma:2004zq,Azuma:2004ie}.
From the string-theoretic point of view,
this can be understood as the Myers effect \cite{Myers:1999ps},
where D-instantons are polarized into a D1-brane with an $S^2$ worldvolume
in the presence of a
three-form flux~\cite{Hollowood:2002ax};
hence the name of the model.

Another important aspect of the polarized IKKT matrix model
is that the partition function can be calculated exactly \cite{Komatsu:2024ydh}
by the SUSY localization method \cite{Pestun:2007rz}
analogously to
the BMN model \cite{Asano:2014vba}.
In particular, 
it was found that the partition function diverges
as $ Z \sim \Omega^{-2(N - 1)}$ in the $\Omega \to 0$ limit and
does not converge to that of the original IKKT model, which is
known to be finite \cite{Moore:1998et,Krauth:1998xh,Austing:2001pk,Kazakov:1998ji}.
Also it was found that there is
a
phase transition at $\Omega \sim \mathcal{O}(N^{-1/2})$
by investigating the model obtained
by the localization method numerically \cite{Hartnoll:2025ecj}.

In this paper, we perform Monte Carlo simulations of the polarized IKKT model
in the case of matrix size $N=2$ for a wide region of $\Omega$.
In particular, we capture
the competing saddle points reliably by using the parallel tempering,
which was not done in the previous
preliminary
studies \cite{Kumar:2022giw,Kumar:2023nya}.
This plays a crucial role in
reproducing the partition function obtained by the localization
method precisely.
Furthermore, we
probe directly the spacetime structure emergent
from the dominant matrix configurations,
which is not accessible
by the localization method.
Thus our results provide a complete understanding of
the nature of the transition at intermediate $\Omega$
as well as the singularity
in the $\Omega \rightarrow 0$ limit.



\textit{The polarized IKKT matrix model---}The action of the Euclidean IKKT model
is given by
\begin{align}
\label{S_IKKT}
S_{\text{IKKT}} &=
\text{tr}\left\{  -\frac{1}{4}\left[A_\mu,A_\nu\right]^2
-\frac{i}{2} \Psi_\alpha ( {\cal C} \Gamma^\mu)_{\alpha\beta} [A_\mu,\Psi_\beta]  \right\} \,  ,
\end{align}
where
$A_\mu$ and $\Psi_{\alpha}$  are
10 bosonic and 16 fermionic $N\times N$ traceless hermitian matrices,
which transform as a vector and a Majorana-Weyl spinor in 10D.
$\Gamma^\mu$ are the 10 dimensional Euclidean gamma matrices
after Weyl projection and ${\cal C}$ is the charge conjugation matrix.
The polarized IKKT model is obtained by
adding the terms \cite{Bonelli:2002mb}
\begin{align}
\label{S_omega}
      S_\Omega &= \text{tr}\left\{  \frac{\Omega^2}{4^3}
      \left(  3A_a^2+A_I^2\right)+i \, \Omega[A_1, A_2] A_3  \nonumber \right. \\
&  \qquad   \ \left.  -\frac{\Omega}{8}
      \Psi_\alpha ({\cal C}\Gamma^{123})_{\alpha\beta} \Psi_\beta   \right\}
\end{align}
corresponding to the SUSY deformation, 
where $a=1,2,3$ and $I=4,\cdots,10$ denote the polarized and
unpolarized directions, respectively,
and $\Gamma^{123} \equiv \Gamma^1(\Gamma^2)^\dag\Gamma^3$.
%
The sign of $\Omega$ is irrelevant since it can be absorbed by
$A_\mu \rightarrow - A_\mu$ and $\Psi_\alpha \rightarrow i \Psi_\alpha$.

At $\Omega\gg1$,
the path integral is dominated by
the classical solution
\begin{eqnarray}
  A_a=\frac{3}{8}\Omega\,J_a\,,\quad A_I=0\,,\quad\Psi_\alpha=0 \, ,
  \label{eq:fuzzy-sphere}
\end{eqnarray}
where $J_a$ is the $N$-dimensional irreducible representation
of $\mathfrak{su}(2)$,
which represents a single fuzzy sphere
in the polarized directions.

In the $\Omega \rightarrow 0$ limit,
the polarized IKKT model has a diverging partition function,
and it does not reduce to 
the original IKKT model as already mentioned.
It was recently pointed out \cite{Komatsu:2024ydh} 
that
this singularity
is due to the commuting matrix configurations.

In the original IKKT model, the classical solutions are indeed given
by commuting matrices, which can be represented by diagonal matrices
\begin{align}
  A_\mu = {\rm diag} (x_\mu^{(1)} , \cdots , x_\mu^{(N)}) \, ,
  \quad \mbox{~where~} \sum_{i=1}^N  x_\mu^{(i)} = 0 \, ,
\label{cl-sol-original}
\end{align}
using ${\rm SU}(N)$ symmetry. By integrating out the off-diagonal components
at the one-loop level, one may attempt to obtain a low-energy effective
theory \cite{Aoki:1998bq}, which is valid when all the diagonal components are
separated from each other.
Due to supersymmetry, the one-loop contributions from
the bosonic and fermionic off-diagonal components cancel each other.
However, the fermionic diagonal components
do not have quadratic terms,
which makes the integration over them nontrivial.

The situation simplifies for $\Omega \neq 0$ since
the $\mathcal{O}(\Omega)$ fermionic mass term in \eqref{S_omega}
induces the quadratic terms of the fermionic diagonal components
and one can integrate them out trivially.
Thus the one-loop effective theory
becomes \cite{Komatsu:2024ydh} 
\begin{align}
\label{Z_1loop}
Z_{\text{1-loop}}
&= \Omega^{8(N-1)} \int dx \exp\left\{
-\frac{\Omega^2}{2^7}\left(3(x_a^{(i)})^2 + (x_I^{(i)})^2\right)\right\}\, .
\end{align}

\textit{Saddle-point equation---}In order to discuss the matrix configurations that dominate
the path integral in the polarized IKKT model,
let us first integrate out the fermionic matrices
to obtain the bosonic integral
\begin{eqnarray}
\label{partition_Z}
Z(\Omega)=\int dA \,\text{Pf}(M(A))\,e^{-S_{\rm b}(A)}
  = \int dA \, e^{-S_{\rm eff}(A)} ,
\end{eqnarray}
where $S_{\rm b}$ is the bosonic part of the action and $\text{Pf}(M)$
denotes the Pfaffian of the antisymmetric matrix $M(A)$ that appears as the kernel
in the fermionic part of the action.
The effective action $S_{\rm eff}(A)$ in \eqref{partition_Z}
is given by
\begin{align}
  S_{\rm eff}(A) = S_{\rm b}(A) -
  \log \text{Pf}(M(A)) \, ,
  \label{eff-action}
\end{align}
and the saddle-point equation reads
\begin{eqnarray}
  0=\frac{d S_{\rm eff}}{dA} 
  =\frac{dS_{\rm b}}{dA}-\frac{1}{2}\text{Tr}\left( M^{-1}\frac{dM}{dA}  \right) \, .
  \label{eq:spa}
\end{eqnarray}

In what follows,
we discuss the solutions to \eqref{eq:spa}
in the $N=2$ case,
for which the Pfaffian is real and positive semi-definite.
This makes the saddle points real
and enables us
to perform Monte Carlo simulations of the model \eqref{partition_Z}
without suffering from the sign problem,
which is not the case for larger $N$.


\textit{Fuzzy sphere saddle---}Let us first consider the original IKKT
model ($\Omega=0$). For the present $N=2$ case, one can
use the ${\rm SO}(10) \times {\rm SU}(2)$ symmetry
to bring an arbitrary matrix configuration into the form
\begin{eqnarray}
  A_a=x_a \frac{\sigma_a}{2}\,,\quad
  A_I=0 \,  ,
  \label{ansatz-ikkt}
\end{eqnarray}
where $\sigma_a$ ($a=1,2,3$) are the Pauli matrices.
Plugging this into
\eqref{eff-action} for $\Omega=0$,
we get
\begin{align}
\label{eq:Scompact}
S &= \frac{1}{4}B  - 8\log (2C) \, ,
\end{align}
where \( B = (x_1 x_2)^2 + (x_2  x_3)^2 + (x_3 x_1)^2 \) and \( C = x_1 x_2 x_3 \).
 The saddle point is obtained as
\begin{eqnarray}
\label{IKKT_saddle}
    x_1 = x_2 = x_3 = 2^{3/4}\, ,
\end{eqnarray}
up to the sign flip.
This is different from the classical solution \eqref{cl-sol-original}
due to the Pfaffian, which takes into account
the full quantum effects of the fermionic matrices.

For $\Omega\neq0$,
we take \eqref{ansatz-ikkt}
with $x_1 =x_2 = x_3 \equiv x$  as an ansatz
in view of \eqref{IKKT_saddle}.
The effective action \eqref{eff-action} becomes
\begin{align}
  S_{\rm eff} &=
  \frac{3}{4} x^4
  + \frac{9 \Omega^2}{2^7} x^2
  - \frac{\Omega}{2} x^3
  - \log \left(2 x^3 + \frac{3\Omega}{4}x^2 - \frac{\Omega^3}{64}\right)^8 \, .
\end{align}
The saddle-point equation
admits a closed-form solution,
which
shall be given elsewhere \cite{longpaper}.  
Here we present the asymptotic behaviors of
the relevant saddle point identified in our simulation
\begin{eqnarray}
  \label{eq:asymptotic-behaviors}
x=
\begin{cases}
  2^{3/4}+ \frac{3}{32} \Omega+ \mathcal{O}\!\bigl(\Omega^{2}\bigr)
  & \mbox{for~}\Omega \ll 1   \, , \\
  \frac{3}{8}\Omega +\mathcal{O}\!\bigl(\Omega^{-2}\bigr)
   &  \mbox{for~}\Omega\gg1 \, .
\end{cases}
\end{eqnarray}
This shows that in the present case of $N=2$,
the unique saddle point \eqref{IKKT_saddle}
of the original model
is smoothly connected to the dominant solution
\eqref{eq:fuzzy-sphere} at $\Omega \gg 1$.




\textit{Commuting saddle---}For $\Omega \ll 1$,
the dominant saddle point is expected to be commuting matrices
from the discussion below \eqref{cl-sol-original}.
Using the ${\rm SU}(2)\times {\rm SO}(3)\times {\rm SO}(7)$ symmetry,
the general form of the commuting matrices
can be put into the form
\begin{eqnarray}
  A_3=x\frac{\sigma_3}{2}\,,\quad A_{10}=y\frac{\sigma_3}{2}\,,\quad
  A_\mu=0\,\text{(otherwise)}\, .
\end{eqnarray}
Plugging this into \eqref{eff-action}, we get
\begin{align}
\label{diverging S}
S &= \frac{\Omega^2}{2^7}(3x^2 + y^2) - \log\left(\frac{\Omega^8}{2^{16}} E
    \right)\,,
\end{align}
where $E=\left\{ (x^2 + y^2)^2 + {\Omega^2}(-x^2 + y^2) /{2^3}
        + {\Omega^4}/{2^8}\right\}^4 $.

The corresponding saddle points are given by 
\begin{eqnarray}
\label{diverging saddle}
\begin{array}{lll}
\mbox{(i)} & x= \frac{1}{4\Omega}\sqrt{\frac{2^{14}}{3}+\Omega^4} \, , &  y= 0  \, , \\
\mbox{(ii)} & x=0 \, ,  &  y = \frac{1}{4\Omega}\sqrt{2^{14}-\Omega^4}  \, ,
\end{array}
\end{eqnarray}
up to the sign flip,
both of which diverge as $\mathcal{O}(1/\Omega)$ for $\Omega \rightarrow 0 $.
In fact, the effective action \eqref{diverging S} becomes small on
the ellipse in the $xy$ plane including these saddle points \eqref{diverging saddle},
along which the maximum and minimum 
are given by (i) and (ii), respectively.
On the other hand, the fluctuations away from this ellipse
are not suppressed at all.
A better description in this regime is given by
the one-loop effective theory \eqref{Z_1loop} obtained
by treating bosonic and fermionic variables on equal footing.

\begin{figure}[t]
\includegraphics[scale=0.53]{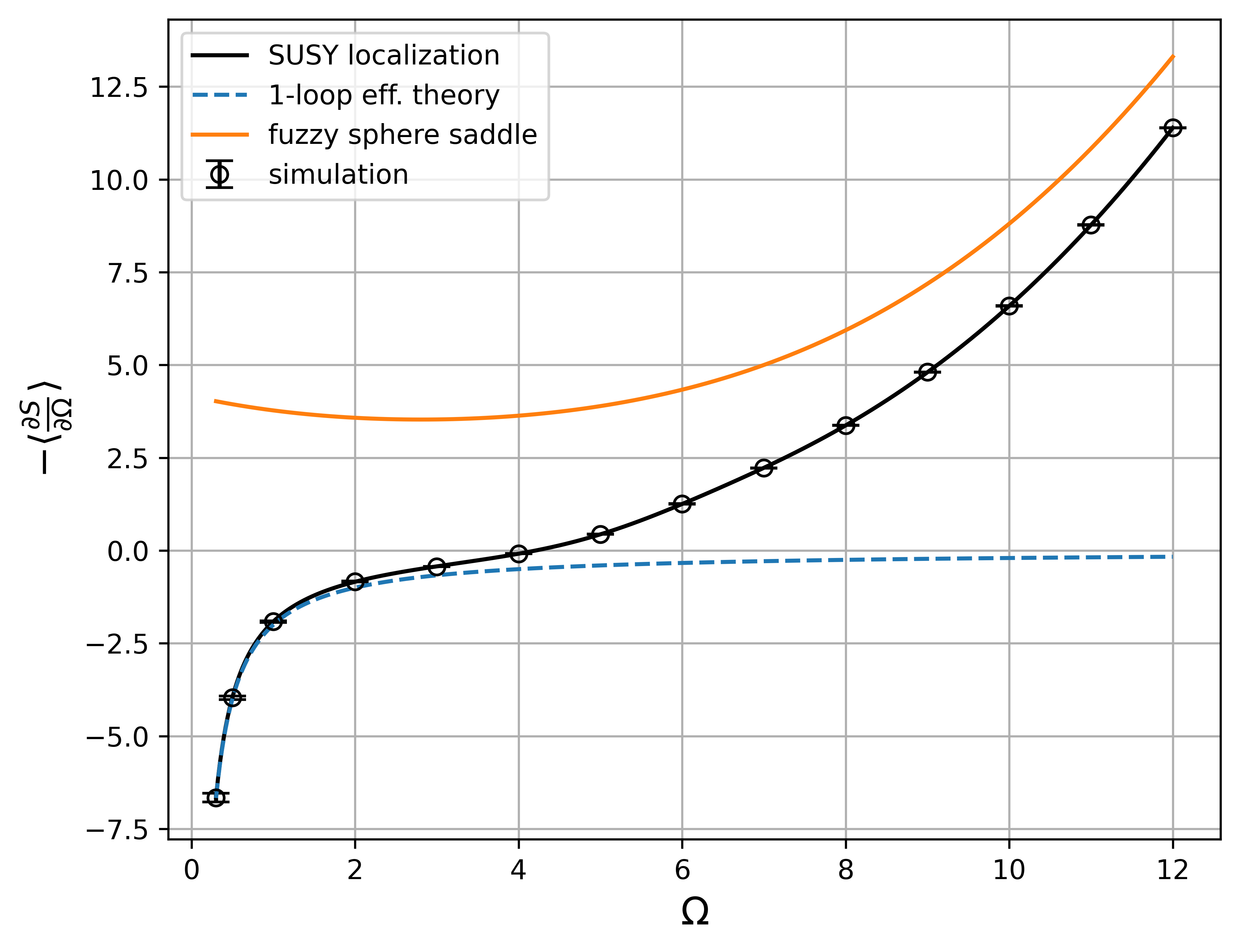}
  \caption{The derivative of the partition function
    $d\log Z(\Omega)/d\Omega = -\langle \partial S/\partial\Omega \rangle$
    obtained by our simulations is plotted against $\Omega$ (black circles).
    The black line represents the result obtained by
    the SUSY localization method.
    The orange line represents the result obtained
    for the fuzzy sphere saddle, whereas
    the blue dashed line represents the result obtained
    by the one-loop effective theory.}
  \label{fig:dSdO}
\end{figure}

\textit{Monte Carlo result---}Let us first
present our results for the partition function.
More precisely, we present the derivative
$d\log Z(\Omega)/d\Omega = -\langle \partial S/\partial\Omega \rangle$,
which is directly accessible by Monte Carlo simulations
as an expectation value.
In Fig.~\ref{fig:dSdO}, we plot
our results for $0.3 \le \Omega \le 12$,
which are in complete agreement with
the results obtained by the SUSY localization method \cite{Komatsu:2024ydh}.

As $\Omega$ increases, we find that the partition function
approaches 
the prediction from the fuzzy sphere saddle,
whereas for $\Omega \rightarrow 0$,
it approaches the result obtained from
the one-loop effective theory \eqref{Z_1loop}.
In particular, we note that $d\log Z(\Omega)/d\Omega \sim - 2/\Omega$
as $\Omega \rightarrow 0$, which shows that $Z(\Omega) \sim \Omega^{-2}$ as
expected for $N=2$.
Note also that the derivative $d\log Z(\Omega)/d\Omega$ changes its sign
at $\Omega \sim 4$, where the dominant configurations are expected to switch
from commuting matrices to the fuzzy sphere saddle.



\begin{figure}[t]
  \centering
  \includegraphics[width=0.48\textwidth]{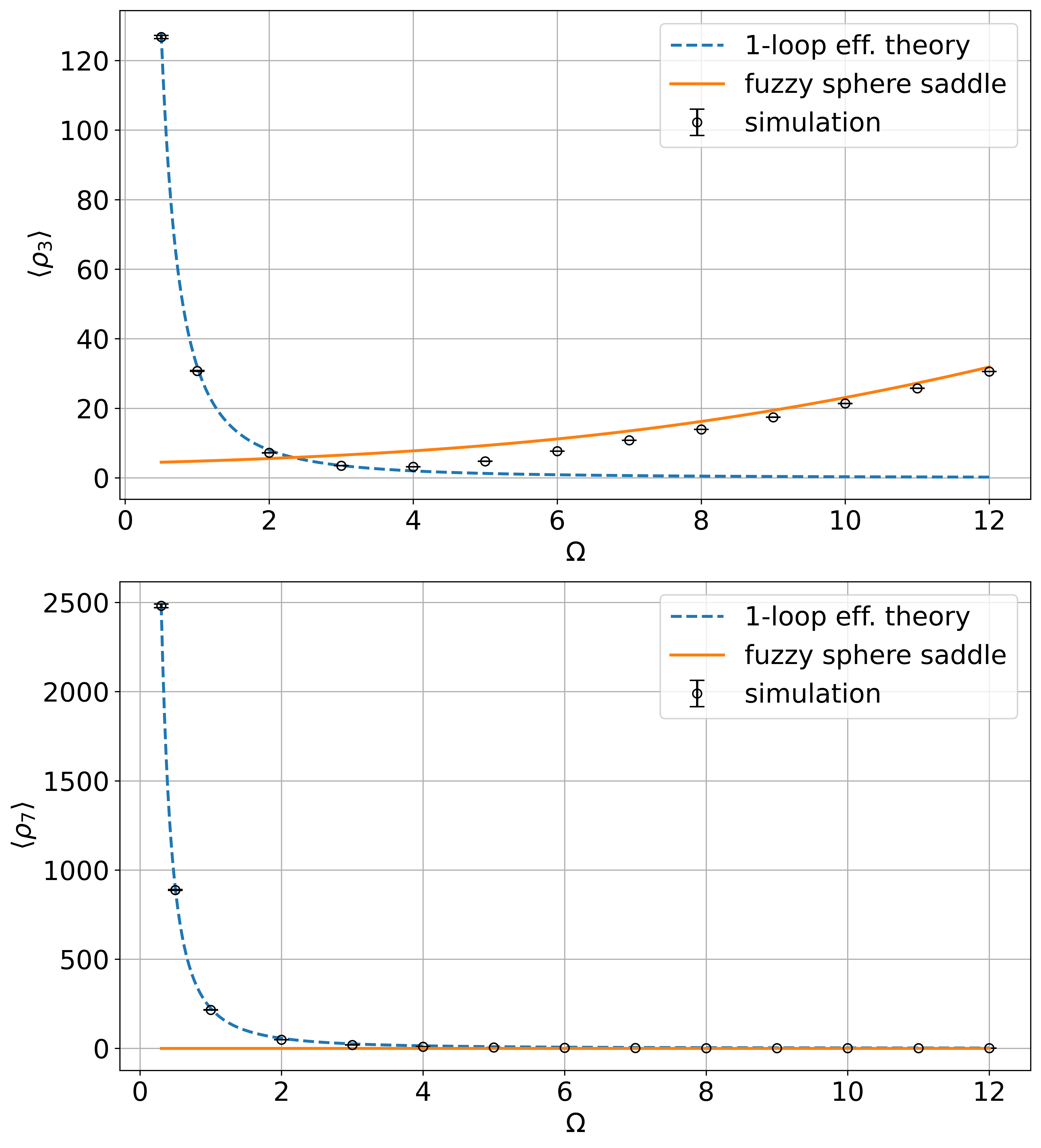}
  \caption{The extent of spacetime $R_3$ (Top) and $R_7$ (Bottom)
    in the polarized and unpolarized directions obtained by simulations are plotted against
    $\Omega$ (black circles).
    The orange line represents the result obtained
    for the fuzzy sphere saddle, whereas
    the blue dashed line represents the result obtained
    by the one-loop effective theory.}
  \label{fig:R3_R7_combined}
\end{figure}

In order to probe the spacetime structure emergent in the IKKT matrix model,
the ``extent of spacetime''
$R^2=\langle \text{tr}\,(A_\mu)^2 \rangle$ \cite{Hotta:1998en}
has been commonly calculated.
In the polarized IKKT model, it is useful to define
\begin{eqnarray}
\label{R3R7 def}
\rho_{3}=\text{tr}\,(A_{a})^2  \,,\quad
\rho_{7}= \text{tr}\,(A_{I})^2  \, 
\end{eqnarray}
separately for the polarized $a=1,2,3$ and unpolarized $I=4,\cdots , 10$ directions.
From the viewpoint of the effective theory of D instantons,
\eqref{R3R7 def} show how their distribution is affected by the background 3-form flux.




In Fig.~\ref{fig:R3_R7_combined}, we plot $\braket{\rho_3}$ and $\braket{\rho_7}$ against $\Omega$.
At large $\Omega$, we find that $\braket{\rho_3}$ grows quadratically with $\Omega$
and $\braket{\rho_7}$ tends to vanish,
as expected from
the fuzzy sphere saddle.
As $\Omega \rightarrow 0$,
both $\braket{\rho_3}$ and $\braket{\rho_7}$ exhibit rapid growth,
which agrees precisely with the predictions from the one-loop effective theory \eqref{Z_1loop}.
This further confirms that the partition function is dominated by
commuting matrix configurations in this regime.

\begin{figure}[t]
    \includegraphics[width=0.48\textwidth]{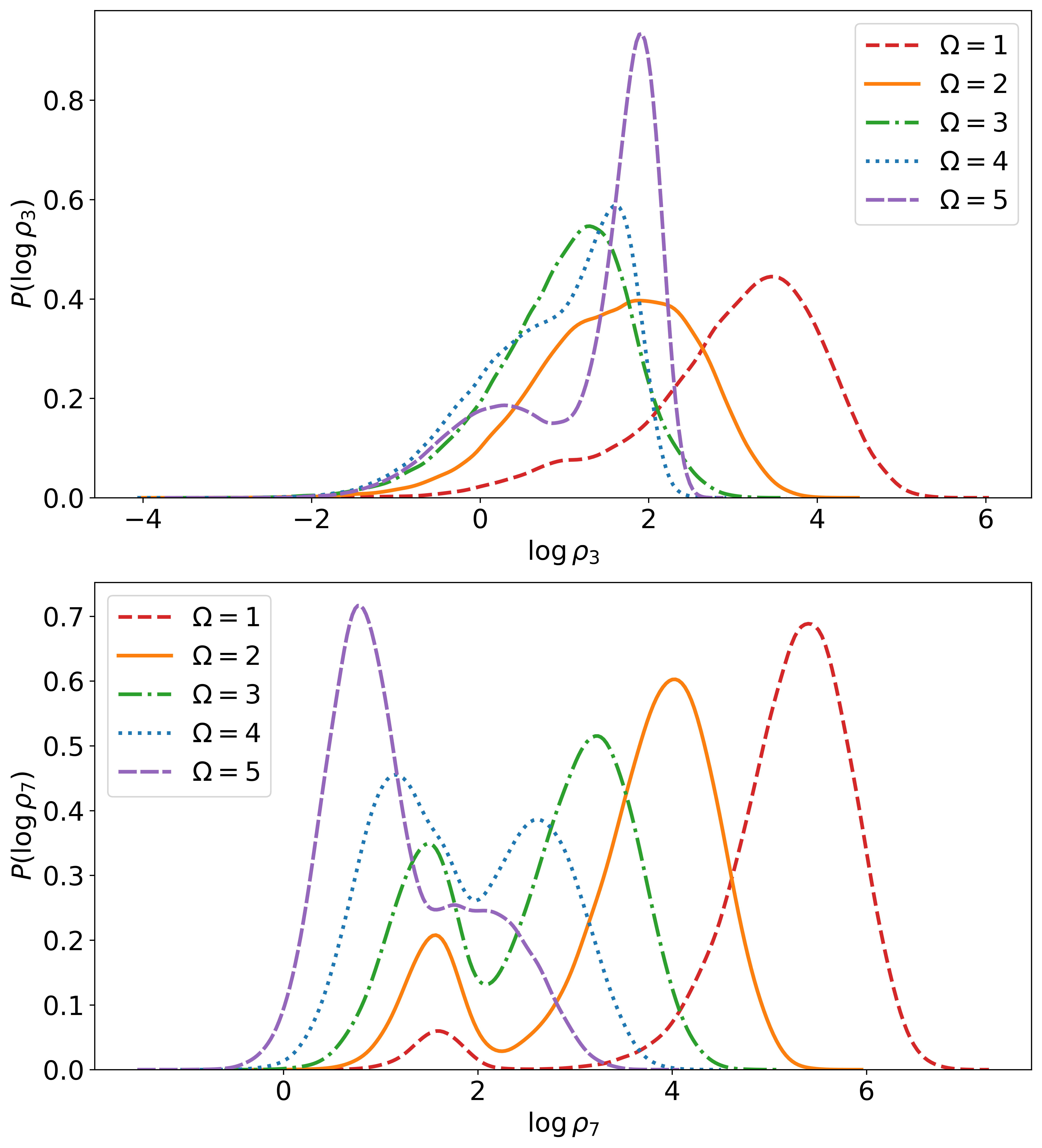}
    \caption{Histogram of the quantities
      $\log \rho_3$ (Top) and $\log \rho_7$ (Bottom),
      which represent the extent of spacetime
      in the polarized and unpolarized directions,
      are shown for $\Omega=1,2, \cdots, 5$.
}
\label{fig:hist_combined}
\end{figure}



Finally, let us see more in detail what happens 
in the intermediate regime at $\Omega \sim 4$,
where a change in the dominant configurations is anticipated.
In Fig.~\ref{fig:hist_combined}, we plot the histogram
of $\log \rho_3$ and $\log \rho_7$, where
we see clear double-peak structures showing up at some values of $\Omega$.
Even at \(\Omega = 1\), the \( \rho_7\) distribution has two peaks,
the right one corresponding to the commuting matrices
and the left one corresponding to the fuzzy sphere saddle,
which becomes the unique
saddle of the original model as $\Omega \rightarrow 0$.
We find that the right peak is by far dominating, which explains clearly why
one cannot retrieve the original IKKT model in the $\Omega \rightarrow 0$ limit.
As \(\Omega\) increases, the right peak becomes smaller and comes closer to the left peak.
In particular, at $\Omega \sim 4$, the two peaks become comparable
as expected from the behavior of the partition function.
At $\Omega \sim 5$, the two peaks merge since \(\rho_7\) becomes small
for both contributions.

On the other hand, the \( \rho_3\) distribution starts to have two peaks at 
\(\Omega = 5\), 
the right one corresponding to the fuzzy sphere saddle
and the left one corresponding to the commuting matrices.
As $\Omega$ increases further, the right peak shifts to the right and
becomes more dominant.

While our results demonstrate a clear transition of the dominant configurations
at $\Omega \sim 4 $, the observables are found to be continuous
as we have seen in Figs.~\ref{fig:dSdO} and \ref{fig:R3_R7_combined}
unlike the results of the localization method obtained for $N=40$ \cite{Hartnoll:2025ecj}.
We consider that this is simply because of the chosen matrix size $N=2$.
At larger $N$, the dominance of one of the peaks occurs more rapidly as one crosses
the critical $\Omega$ since  the associated free energy is $\mathcal{O}(N^2)$.
In fact, the calculation for small $N$ is more tricky since one has to sample
both peaks with the correct weight for a wide region of $\Omega$,
where the two peaks are actually very separated.
This is made possible by
using a sophisticated parallel tempering HMC algorithm
as we discuss in a separate paper \cite{longpaper}.
The agreement with the result of the localization method in
Fig.~\ref{fig:dSdO} is achieved only with such calculations.



\textit{Discussions---}In this letter, we have performed Monte Carlo simulations
of the polarized IKKT matrix
model \cite{Bonelli:2002mb},
which has attracted a lot of attention
recently in the context of gauge-gravity
duality \cite{Hartnoll:2024csr,Komatsu:2024bop,Komatsu:2024ydh}.
In particular, by focusing on
the simplest $N=2$ case,
we were able to identify all the saddle points that contribute to the path integral.
While the validity of our simulations
is confirmed by the precise agreement with the result
of the localization method, the observables such as $\rho_3$ and $\rho_7$, which are
not accessible by the localization method, tell us clearly the spacetime structure of the
dominant configurations depending on $\Omega$.
In particular, this clarified the geometric
nature of the transition at intermediate $\Omega$.



The fact that one cannot retrieve the original IKKT model
in the $\Omega\rightarrow 0$ limit
looks surprising at first sight.
Here we point out that this effect is actually quite generic,
and there is no need for SUSY or even fermions.
For instance,
let us consider a simple one-variable integral
\begin{equation}
  Z = \int_{-\infty}^{\infty} \mathrm dx\; e ^{-V(x)} \, ,
  \label{eq:Zdef}
\end{equation}
with a polynomial ``potential''
\begin{equation}
V(x) =   x^{2} - \Omega^{2}x^{4} + a \, \Omega^{4} x^{6} \, , 
  \label{eq:V-x}
\end{equation}
where $a > 0$ and $\Omega \in \bbR$.
While this
is an innocent-looking
deformation
of the Gaussian integral ($\Omega=0$),
the integral actually diverges in the $\Omega \rightarrow 0$ limit
for $a<1/4$.

To see that, we rewrite the integral \eqref{eq:Zdef}
in terms of the rescaled variable $y \equiv \Omega \, x$ as
\begin{equation}
  Z=\frac{1}{\Omega}\int_{-\infty}^{\infty}\mathrm dy\, e^{-\tilde{V}(y) / \Omega^2}
  \, , \quad
  \tilde{V}(y) =y^2\bigl(1-y^{2}+ a \, y^{4} \bigr) \, .
  \label{eq:Zscaled}
\end{equation}
For $a<1/3$, one obtains a new minimum of $\tilde{V}(y)$
at $y=y_0$ [with $(y_0)^2=(1+\sqrt{1-3a})/3a$]
in addition to the trivial one at $y=0$.
And for $a < 1/4$, the new minimum gives
$\tilde{V}(y_0)<0$, hence the divergence as $\Omega \rightarrow 0$.
In terms of the original integral, this happens because
the new minimum at $x= x_0 \equiv y_0/\Omega$ becomes infinitely deep
$V(x_0) = - c/\Omega^2$ with $c=|\tilde{V}(y_0)|>0$.
Note also that the new minimum at $x=  y_0/\Omega$
is pushed away to infinity as $\Omega \rightarrow 0$, which ensures the consistency
with the fact that there is no such a minimum at $\Omega=0$.
In the polarized IKKT model, the role of the new minimum is played by the
commuting matrices.




This new insight into the singularity in $\Omega\rightarrow 0$
may have broad implications in physics.
For instance, if one applies it to quantum field theory
(or to some quantum system),
it implies that an infinitesimal deformation of the theory may lead to
the emergence of a totally different vacuum (or a totally different ground state).
In the case of superstring theory, it is known that there are tremendously
many vacua, which are perturbatively stable, and each vacuum corresponds
to a spacetime with different dimensionality accommodating quantum fields
with different gauge symmetry.
This is the situation that is commonly referred to as the string landscape.
The IKKT matrix model is supposed
to describe each vacuum in the string landscape as saddle points.
However, a small deformation may lead to a new saddle point,
which is actually dominant.

In this regard, let us recall that the polarized IKKT model is a deformation
of the \emph{Euclidean} IKKT
model \cite{Nishimura:2011xy, Anagnostopoulos:2020xai,Anagnostopoulos:2022dak}
that can be obtained
by applying a Wick rotation $A_0 = i A_{10}$ to the \emph{Lorentzian}
IKKT model.
In fact, in the Lorentzian model,
by adding a Lorentz invariant mass term
$S_{\rm m}= - \gamma {\rm tr} (A_\mu A^\mu)$ to the action,
one can obtain new saddle points, which represent expanding spacetime
\cite{Kim:2011ts,Kim:2012mw,Steinacker:2017vqw,Steinacker:2017bhb,Hatakeyama:2019jyw,Sperling:2019xar,Steinacker:2021yxt}.
It would therefore be interesting to investigate
the SUSY mass deformation of the Lorentzian model analogous to the polarized IKKT model.


Recently it was recognized that the Lorentzian model has a diverging partition function
due to the noncompact Lorentz symmetry \cite{Asano:2024def}.
This led to a proposal of
a well-defined model obtained by
fixing the Lorentz symmetry by the Faddeev-Popov procedure \cite{Asano:2024def}.
Monte Carlo simulations of 
the Lorentzian model defined in this way
were performed with the Lorentz invariant mass term
omitting the fermionic matrices \cite{Chou:2025moy}.
We are currently trying to extend this work to the SUSY deformed model
including the fermionic matrices.

\textit{Acknowledgement---}The authors would like to thank Yuhma Asano,
Tin-Long Chau,
Heng-Yu Chen, Yu-An Chen,
Franz Ciceri,
Sean A. Hartnoll, Pei-Ming Ho, Masazumi Honda, Hikaru Kawai,
Shota Komatsu, Henry Liao, Jun Liu, Ashutosh Tripathi
and Asato Tsuchiya for valuable discussions.
Simulations were carried out
on the KEK Central Computing System
and a PC cluster at KEK.
C.-Y. C. was supported by JST SPRING, Japan Grant Number JPMJSP2104.

\bibliography{div-saddle}

\end{document}